\documentclass[aps,prl,reprint]{revtex4-1}
\usepackage[utf8]{inputenc}
\usepackage{graphicx}
\usepackage{blindtext}
\usepackage{calrsfs}
\usepackage[colorlinks=true, allcolors=blue]{hyperref}
\usepackage{mathtools}
\usepackage{amsmath}
\DeclareMathAlphabet{\pazocal}{OMS}{zplm}{m}{n}

\begin{document}


\title{Dual-mode phonon dynamics and lasing in optomechanical cavities}

\author{Raúl Ortiz}
\email{rortfer@ntc.upv.es}
\affiliation{Nanophotonics Technology Center, Universitat Politècnica de València, Building 8F, 46022 Valencia, Spain}

\author{Carlos Mas Arabí}
\author{Carles Milián}
\affiliation{Institut Universitari de Matemàtica Pura i Aplicada, Universitat Politècnica de València, 46022 Valencia, Spain}

\author{Alejandro Martínez}
\affiliation{Nanophotonics Technology Center, Universitat Politècnica de València, Building 8F, 46022 Valencia, Spain}

\date{\today}

\begin{abstract}
We use the full nonlinear bifurcation theory as a powerful methodology to thoroughly classify and predict the phonon lasing dynamics in optomechanical cavities. We exemplify its scope in the very relevant and so far vaguely explored dynamics of dual-mode phonon lasing when two independent mechanical modes are directly coupled to one optical field. We uncover a plethora of different lasing regimes in a fixed and realistic cavity geometry, including single- and dual-mode lasing, perfect synchronization of different mechanical modes, and quasiperiodic time crystals. All dynamical regimes are unambiguously associated with bifurcations of different natures, which may exhibit both super and sub-critical natures characterizing the often hysteretic behavior of the systems when they are externally driven by a time-varying laser source. Our results, generalizable to any other optomechanical system, open unprecedented pathways to understand and control the formation and dynamics of optomechanical frequency combs.

\end{abstract}

\maketitle

Optomechanics (OM) studies the interaction between light and mechanical waves in both quantum and classical realms, ranging from individual photons to classical light beams as well as from molecular phonons to macroscopic elastic waves \cite{OM_History1,OM_History2,OM_History3,OM_History4,OM_History5}. When confined in cavities, light and mechanical waves can interact, affecting the dynamics of each other via the so-called OM backaction \cite{CavOpt}. Under a regime of blue-detuned laser driving, OM backaction yields phonon lasing \cite{CavOpt}, which in a cascaded effect results in the formation of highly-coherent OM frequency combs with spacing equal to the mechanical resonance (typically below 10 GHz, depending on the oscillator size). Such OM frequency combs extend over a bandwidth much narrower than purely optical combs (reaching even more than one octave \cite{Octave1,Octave2,Octave3}), but the fact that the line spacing is much smaller than in the latter case, makes them very appealing for applications such as all-optical frequency conversion \cite{5G}, sensing \cite{Sensing1} or particle levitation \cite{Levitate}.\\
\indent
Formal developments in the OM field usually rely on analytical methods applied to linearized models that typically predict the onset of several phenomena of great importance such as the first phonon lasing threshold, well known to occur at a Hopf bifurcation \cite{Hopf_paper}, OM cooling \cite{OMCooling}, normal mode splitting \cite{OMSplitting}, amplification \cite{OMAmplification} or four-wave mixing \cite{OM4WaveMixing}. Beyond the perturbative limits, the study of nonlinear phenomena at large pumps to reach synchronization, dual-mode lasing, quasiperiodic states, and chaos requires fully numerical investigations that are, however, often based on blind \textit{brute force} propagation simulations failing to provide an overall view on the system within reasonable timescales. Hence, the lack of a well-established methodology in the field precludes timely development of the understanding of complex OM cavity systems.



In this Letter, we unveil the dynamics of a multimode OM cavity through rigorous bifurcation analysis and numerical continuation. This methodology yields a precise and comprehensive description of the system's intrinsic nonlinear nature and provides the shape and stability of the resulting OM frequency combs. Commonly used in nonlinear science, these mathematical tools have been successful in applications ranging from optical frequency combs \cite{Examples_1,Examples_2} to biological systems \cite{Examples_3} and vegetation modeling \cite{Examples_4,Examples_5}, yet they remain largely unexplored in the context of OM.

So far, the simultaneous parametric amplification of multiple mechanical oscillators coupled to an optical mode was often limited by mode competition, resulting in the amplification of a single mode and damping the rest \cite{Lasing_Competition1,Lasing_Competition2,General_Coup2}. This effect typically arises unless the mechanical resonators have notably different frequencies \cite{Lasing_Low_High0, Lasing_Low_High1,Dual_Lasing_Exp} or the optical driving is modulated. The simultaneous and coherent emission of two phonons with similar frequencies without modulating the optical driving has remained overlooked, as it arises in a regime where the system exhibits strong nonlinear behavior. In our work, we apply bifurcation analysis and numerical continuation to an OM cavity comprising two independent mechanical modes. Remarkably, we identify a configuration where phonon emission occurs at two distinct but similar frequencies.

\begin{figure*}[t]
    \centering
    \includegraphics[width=0.8\textwidth]{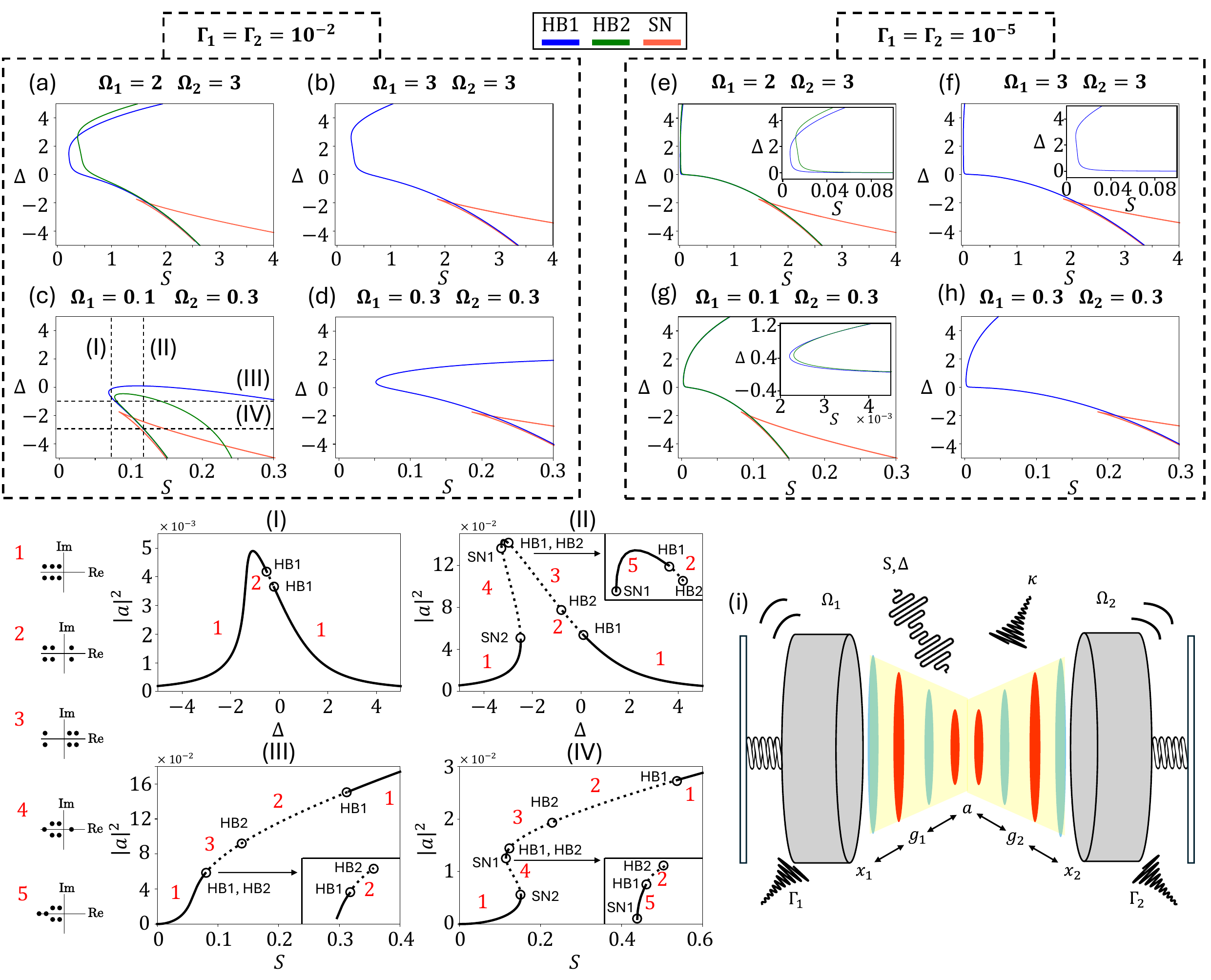}
    \caption{Figures (a)-(h) show the first (HB1, blue) and second (HB2, green) Hopf bifurcations, along with saddle-node bifurcations (SN, orange), for different mechanical frequency configurations and dissipation rates. The left (right) column corresponds to high (low) mechanical dissipation rates, $\Gamma_1 = \Gamma_2 = 10^{-2}$ ($\Gamma_1 = \Gamma_2 = 10^{-5}$). Figures (I)-(IV) show steady-state solutions of $|a|^2$ as a function of detuning $\Delta$ ((I),(II)) and $S$ ((III),(IV)) for selected parameter cuts in (c), illustrating bistability and Hopf bifurcations. Red-detuned cases (II),(IV) exhibit three steady-state branches separated by SN bifurcations, while blue-detuned cases (I),(III) transition directly to oscillations at lower thresholds. Numbers indicate stability regions that correspond to the eigenvalues: 1 and 5 (stable steady-state), 2-3 (periodic solutions), and 4 (instability). Insets in (e)-(g) and (II)-(IV) provide a magnified view of the regions where the curves are most densely packed. Figure (i) depicts an OM cavity where a single optical mode is coupled to two mechanical modes (two movable mirrors).}
    \label{fig:bifurcations}
\end{figure*}

\begin{figure*}[t]
    \centering
    \includegraphics[width=0.8\textwidth]{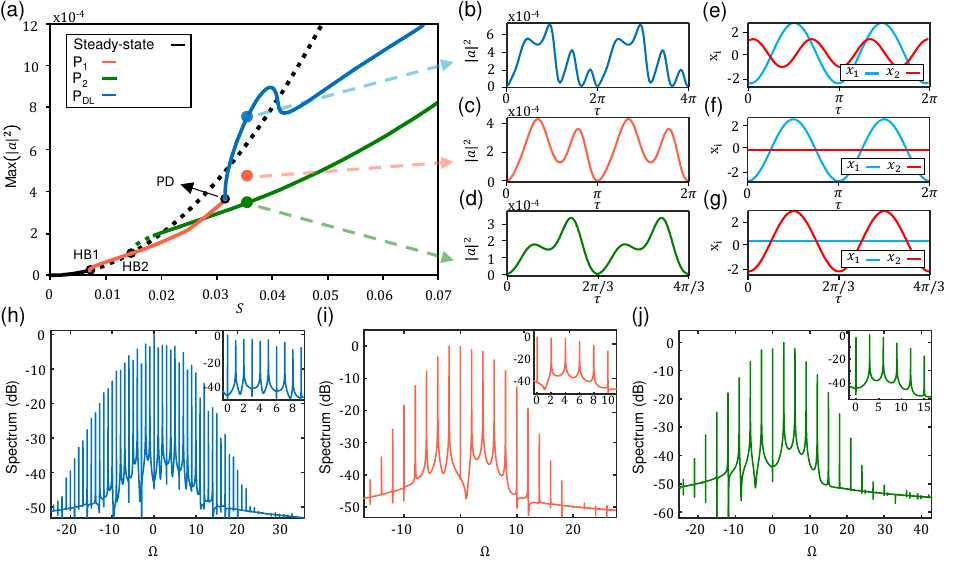}
    \caption{Periodic solution branches for the values $\Omega_{1}=2$, $\Omega_{2}=3$, $\Gamma_{1}=\Gamma_{2}=10^{-5}$ and $\Delta=1$. (a) Periodic oscillation branches ($P_1$ and $P_2$) after two Hopf bifurcations (HB1 and HB2) and dual-mode lasing branch ($P_{DL}$) after a period doubling bifurcation (PD). (b)-(d) Optical intensity profiles over two periods for the branches shown in (a) at $S = 0.035$. (e)-(g) Corresponding mechanical oscillations for the three branches at $S = 0.035$. Periods in (b)-(g) differ as they correspond to distinct solution branches. (h)-(j) Spectra of the optical intensity signals in (b)-(d), respectively, in log scale, highlighting the dominant frequency components. Dashed lines in (a) show unstable solutions.}
    \label{fig:3_Branches}
\end{figure*}

\begin{figure*}[t]
    \centering
    \includegraphics[width=\textwidth]{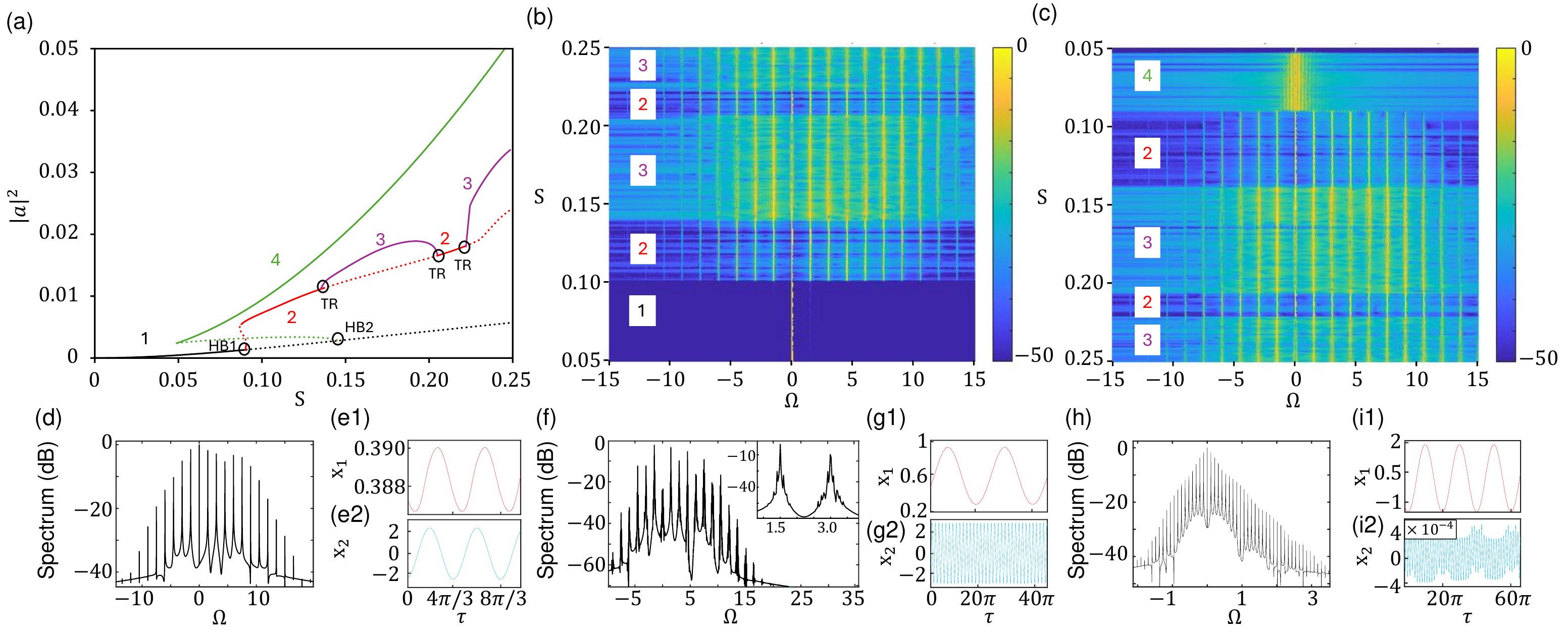}
    \caption{Bifurcation diagram and dynamical behavior of the system for $\Omega_1=0.1$, $\Omega_2=1.5$, $\Gamma_1=\Gamma_2=10^{-3}$, and $\Delta=2$. (a) Bifurcation diagram showing different dynamical regimes labeled 1 to 4. Region 1: steady-state solution. Region 2: synchronization of both mechanical modes to $\Omega_2$. Region 3: quasiperiodic dual-mode lasing originating from a torus bifurcation (TR). Region 4: enhanced oscillation of the second mechanical mode ($\Omega_1$) while the first mode remains damped. (b) Forward sweep of the pump $S$ showing transitions between regions, including the emergence of a dual-mode frequency comb. (c) Backward sweep showing hysteresis and transition into region 4. (d), (f), and (h): Spectra of the optical output for $S=0.11$, $0.16$, and $0.06$, respectively, illustrating synchronized, dual-mode, and single-mode frequency combs. (e), (g), and (i): Corresponding mechanical oscillations for each case.}
    \label{fig:Third_Plot}
\end{figure*}

Our results reveal that dual-mode phonon lasing can emerge in a regime where both mechanical modes oscillate with comparable amplitudes following a period-doubling bifurcation. Notably, we find that the period-doubling bifurcation modifies the frequency comb structure and plays a crucial role in triggering the lasing of the second mechanical mode under specific conditions, unveiling an unexpected pathway to dual-mode phonon lasing. We also identify another distinct dual-mode lasing state in the limit where mechanical frequencies are of different orders of magnitude. Such states exhibit quasiperiodic behavior that appears after a torus bifurcation. Furthermore, we demonstrate that synchronization between the two mechanical modes can occur, even when their natural frequencies are significantly different---an intriguing and nontrivial effect that emerges under specific conditions and is captured by our dynamical analysis.
\indent
\indent
We consider a general optically-driven OM cavity with two movable mirrors, each having its natural mechanical frequency ($\Omega_i$), damping rate ($\Gamma_i$), optical losses ($\kappa$), and OM coupling rate ($g_i$) [see Fig. \ref{fig:bifurcations}(i)]. The normalized equations of motion for the intracavity optical field ($a$) and the mirror displacements ($x_i$) are  \cite{CavOpt}:  


\begin{subequations}
\begin{align}
\frac{d a}{d \tau} &= \left(i\left(\Delta + 2 x_{1} + 2 x_{2}\right) - 1\right) a + S, \\
\frac{d^2 x_{1}}{d \tau^2} &= -\Omega_{1}^2 x_{1} - \Gamma_{1} \frac{d x_{1}}{d \tau} + \left|a\right|^2, \\
\frac{d^2 x_{2}}{d \tau^2} &= -\Omega_{2}^2 x_{2} - \Gamma_{2} \frac{d x_{2}}{d \tau} + \left|a\right|^2,
\end{align}
\label{2OMNorm}
\end{subequations}
where $\Delta$ corresponds to the frequency detuning between the laser and the closest optical resonance, and $S$ is the input photon flux. The explicit relation between these normalized and physical parameters is provided in the Supplementary Material.




%

Single-mode phonon lasing occurs when one of the movable mirrors undergoes self-sustained, coherent mechanical oscillations driven by the OM interaction. Such an effect arises at Hopf bifurcations where the stationary states ($\dot{a}=\dot{x_1}=\dot{x_2}=0$) lose stability and give rise to oscillatory behavior, see Supplementary Material. This transition appears when a pair of complex-conjugate eigenvalues of the Jacobian matrix crosses the imaginary axis.\\
\indent
Figures \ref{fig:bifurcations}(a)-(h) present the first (HB1) and second (HB2) Hopf bifurcations, as well as saddle nodes (SNs), for both high and low mechanical dissipation rates ($\Gamma_i$). These bifurcations are mapped across different mechanical frequency configurations, including degenerate cases. Breaking the degeneracy in $\Gamma_i$ has a negligible effect on the overall structure of the bifurcation diagrams. Notably, increasing mechanical frequencies shifts all bifurcations higher, whereas lower dissipation rates lead to reduced thresholds. Furthermore, SNs --- which indicate bistability --- occur exclusively in the red-detuned regime ($\Delta<0$). In contrast, phonon lasing thresholds HB1 and HB2 are lower in the blue-detuned regime ($\Delta>0$), favoring oscillatory behavior for lower pumps. As seen in Figs. \ref{fig:bifurcations}(a) and (e), by carefully controlling the detuning, we can selectively excite one mechanical mode. This behavior is reminiscent of the mode competition control described in \cite{General_Coup2}, where the lasing mode is controlled via the pump strength $S$. In our case, the detuning serves a similar role in selecting the lasing mode.
Figure \ref{fig:bifurcations}(I) shows a bifurcation diagram as a function of $\Delta$ for a fixed value $S=0.7$, with $\Omega_1=0.1$ and $\Omega_2=0.3$. We identify a pair of HBs. Within the detuning range between these two HBs, the system acts as a phonon laser at $\Omega_1$. The optical amplitude $a$ and the displacement $x_1$ oscillate at a frequency near $\Omega_1$, while $x_2$ remains stationary. To excite phonon lasing at $\Omega_2$, we need to increase $S$. This leads to a new pair of eigenvalues crossing the imaginary axis, resulting in the emergence of another pair of Hopf bifurcations [see Fig. \ref{fig:bifurcations}(II)]. Increasing $S$ also leads to bistability in the red-detuned region ($\Delta<0$), as shown in Fig. \ref{fig:bifurcations}(II). This bistable region appears between two SNs. In addition to these vertical cuts in Fig. \ref{fig:bifurcations}(c), Figs. \ref{fig:bifurcations}(III) and (IV) display horizontal cuts at fixed detuning values, showing the system’s response as a function of the pump strength $S$ revealing a characteristic S-shaped curve.
As a result of this detailed exploration, Fig. \ref{fig:bifurcations} reveals the rare coexistence of two Hopf bifurcations associated with qualitatively different dynamical thresholds (internal modes) which emerge due to the multimodal nature of our system and present intertwined paths which are easily traceable with our methodology. This alone is of great importance for cavity design since the loci of the different HBs determine the system's lasing thresholds and features. Note previous studies typically locate just the one HB threshold \cite{Navarro-Urrios, 2Res_2, 2Res_3}.





Now, we consider a specific and experimentally accessible blue-detuned case whose normalized and SI parameters are presented in Table I in Supplementary Material. 
By performing a parameter sweep of Eqs. \ref{2OMNorm}(a)-(c) over $S$, and obtaining the maximum cavity intensity, $\mathrm{max}(|a|^2)$, we plot Fig. \ref{fig:3_Branches}(a). The parameters of the system are $\Omega_{1}=2$, $\Omega_{2}=3$, $\Gamma_{1}=\Gamma_{2}=10^{-5}$ and $\Delta=1$. The steady-state 
remains stable until the first HB. From this steady-state solution, two HBs emerge, leading to periodic solution branches $P_1$  and $P_2$. These branches correspond to self-sustained oscillations (phonon lasing) of the first and second mechanical modes, respectively. Additionally, a distinct solution, shown in blue, corresponds to a dual-mode lasing branch ($P_{DL}$), where both mechanical modes are simultaneously in the lasing regime. This blue curve emerges from a period-doubling bifurcation (PD) of the $P_1$ curve at $S=0.0314$. The PD bifurcation in $P_1$ generates new frequency components at half of the mechanical frequency $\Omega_1 = 2$, along with its harmonics, resulting in a frequency comb with integer-spaced lines up to approximately 20. This mechanism effectively drives the second mechanical oscillator at its natural frequency, eventually leading to its lasing. As a result, both mechanical modes enter the lasing regime simultaneously, exhibiting comparable intensities.

Figures \ref{fig:3_Branches}(b)-(g) show the time-domain evolution for a fixed pump value $S=0.035$, illustrating the behavior of the three solution branches identified in (a). Figures \ref{fig:3_Branches}(b)-(d) display two periods of the intra-cavity optical intensity for the $P_{DL}$, $P_1$, and $P_2$ branches, respectively. Figures \ref{fig:3_Branches}(e)-(g) show the oscillation amplitudes of the two mechanical modes, $x_{1}$ and $x_{2}$, for each case. In Fig. \ref{fig:3_Branches}(e), which corresponds to the blue dual-mode lasing branch, both mechanical modes oscillate with comparable amplitudes. In Fig. \ref{fig:3_Branches}(f), corresponding to $P_1$, only the first mechanical mode is in the lasing regime, while the amplitude of the second mode remains negligible. Finally, Fig. \ref{fig:3_Branches}(g) corresponds to $P_2$, where only the second mechanical mode is in the lasing regime whilst the first mode remains unamplified.\\
\indent
Figures \ref{fig:3_Branches}(h)-(j) present the spectra of the time signals in Figs. \ref{fig:3_Branches}(b)-(d), showing the combs. Figures \ref{fig:3_Branches}(i) and (j) correspond to $P_1$ and $P_2$, respectively, and show single comb structures generated by the first and second mechanical modes. In each case, the peaks are centered around the fundamental frequencies $\Omega_{1}$ or $\Omega_{2}$, along with their harmonics. In contrast, Fig. \ref{fig:3_Branches}(h), corresponding to $P_{DL}$, reveals multiple peaks associated with both frequencies $\Omega_{1} = 2$ and $\Omega_{2} = 3$, as well as their harmonics and combinations, such as sum and difference frequencies. This structure forms a dual-frequency comb characteristic of the dual-mode lasing regime. Notably, the solutions in $P_2$ at $S=0.035$ lose stability after several periods, as indicated by the dashed line in (a).\\
\indent
Unlike previous studies that stop at Hopf bifurcations \cite{HB_p2_1,HB_p2_2}, we reveal in Fig. \ref{fig:3_Branches} a novel dual-mode phonon lasing regime triggered by a period-doubling bifurcation, resulting in a rich dual-frequency comb. Although hybridization \cite{General_Coup1}, mode competition, and anomalous cooling \cite{General_Coup2} have been reported, phonon lasing in this context remains scarcely explored \cite{General_Lasing,Dual_Lasing_Exp}. Our work fills this gap, offering a clear, dynamic roadmap for future experiments. Furthermore, all parameters used are within experimentally feasible ranges.

To delve deeper into the system's dynamics, Fig. \ref{fig:Third_Plot} presents a representative case with parameters set to $\Omega_1=0.1$, $\Omega_2=1.5$, $\Gamma_1=\Gamma_2=10^{-3}$, and $\Delta=2$. In Fig. \ref{fig:Third_Plot}(a), we observe that the first Hopf bifurcation (HB1) is supercritical, while the second (HB2) is subcritical. The diagram displays distinct dynamical regimes, labeled 1 to 4. Region 1 corresponds to the steady-state solution. In region 2, both mechanical modes synchronize to the higher frequency $\Omega_2=1.5$. Region 3 is associated with a dual-mode lasing branch that emerges from a torus bifurcation, resulting in a quasi-periodic regime where both mechanical modes oscillate at their respective natural frequencies. Finally, region 4 (highlighted in green) corresponds to the enhanced oscillation of the second mechanical mode $\Omega_1=0.1$, while the first mode remains strongly damped. Due to the complex dynamics involved, it is not possible to directly associate each branch with a single mechanical mode oscillation, as was done in the previous case.\\
\indent
Figures \ref{fig:Third_Plot}(b) and (c) show time propagation simulations where the pump parameter $S$ is swept up and down, respectively. In panel (b), as $S$ increases, the system initially resides in the steady-state (region 1) until it crosses HB1, entering region 2. Here, both mechanical modes become highly excited and synchronize to the higher frequency. Notably, the synchronization mechanism forces the lower-frequency mechanical mode to effectively increase its frequency by a factor of 15 —a remarkably large ratio compared to typical synchronization values reported in OM systems \cite{Syn_1,Syn_2,Syn_3,Syn_4,Syn_5,Syn_6}. Around $S=0.14$, a torus bifurcation (TR) occurs, giving rise to a new branch characterized by quasi-periodic oscillations, where both mechanical modes exhibit enhanced amplitudes at their respective natural frequencies, leading to a dual-mode frequency comb. Interestingly, similar torus and period-doubling bifurcations have been reported in single-mode OM systems \cite{HB_PL}, but their manifestation in a multimode setup—as shown here—reveals a significantly richer dynamical behavior. This branch eventually rejoins region 2, where the modes synchronize again. When $S$ reaches approximately $0.23$, a further transition to region 3 occurs, reinstating the dual-mode lasing regime.
\indent
Figure \ref{fig:Third_Plot}(c) reveals the hysteretic nature of the system. When $S$ is decreased from the dual-mode lasing state in region 3, the system retraces its path until $S\approx 0.09$, where it abruptly transitions to region 4. Finally, the system returns to the steady state as the green branch loses stability at $S=0.05$. Panels (d), (f), and (h) show the spectra for representative points: synchronized single-mode comb ($S=0.11$), quasi-periodic dual-mode comb ($S=0.16$), and single-mode lasing of the second mechanical mode ($S=0.06$). In each case, the corresponding mechanical oscillations are also plotted, clearly illustrating the distinct dynamical regimes.

Beyond the specific parameter configurations discussed here, we conducted a comprehensive analysis to assess the stability and robustness of the dynamical regimes under variations of the system parameters. In particular, we performed extensive parameter sweeps, as presented in the Supplementary Material, where we systematically explored the influence of detuning, mechanical losses, and pump power. From these simulations, we identified a well-defined range of detuning values where both mechanical modes exhibit comparable oscillation amplitudes, which is essential for achieving the dual-mode lasing regime without requiring external modulation. Furthermore, we observed that reducing the mechanical losses enhances the oscillation amplitudes of the mechanical modes, while higher losses tend to suppress them, especially at lower input power. Notably, we also found that the dual-mode lasing regime persists and remains robust even at elevated pump powers, sustaining coherent oscillations of both mechanical modes.\\
\indent
Additionally, we investigated the role of the period-doubling (PD) bifurcation in enabling the transition to dual-mode lasing. Specifically, we varied the mechanical frequencies (not shown). We observed that if, after the PD bifurcation of the lasing mode, the newly generated comb lines coincide with the frequency of the second mechanical mode, this interaction destabilizes the original lasing branch. As a result, a new stable branch emerges from the PD bifurcation, corresponding to the dual-mode lasing regime.\\
\indent
Finally, we performed a dedicated sweep around the torus bifurcation to determine the threshold and characteristics of the quasiperiodic dual-mode lasing regime. Altogether, this systematic study provides a parameter map that demonstrates the robustness and versatility of the proposed dynamics.\\
\indent
To summarize, we used a robust methodology for analyzing the bifurcation structure and dynamical behavior of OM cavities supporting two mechanical modes, targeting the emergence of single- and dual-mode phonon lasing. Our bifurcation analysis, based on steady-state solutions and their stability, reveals a rich sequence of transitions—including Hopf, period-doubling, and torus bifurcations—leading to stable self-sustained oscillations, dual-mode lasing with frequency-comb-like features, synchronization between widely separated mechanical frequencies, and hysteretic behavior (see Figs. \ref{fig:3_Branches} and \ref{fig:Third_Plot}). Furthermore, the comprehensive parameter sweep plots shown in Supplementary Material map the full landscape of dynamical regimes, providing a powerful tool to identify and control complex behaviors in multimode OM systems. These results open new avenues for tailoring nonlinear phonon lasing dynamics and guiding future experimental implementations.
\section*{Acknowledgements}
The authors acknowledge funding from “Generalitat Valenciana” (GVA, CIACIF/2021/006), HORIZON EUROPE Framework Programme (MAGNIFIC 101091968), Generalitat Valenciana (CIGE/2023/126) and Agencia Estatal de Investigación (ALLEGRO PID2021-124618NB-C21, and MUSICIAN PCI2022-135003-2, CHISTERA IV Cofund 2021). C.M.A acknowledges funding from the Ministerio de Universidades through the Beatriz Galindo program  (BG22/00025).

\end{document}


\renewcommand{\thefigure}{S\arabic{figure}}

\title{Supplementary Material \\ ``Dual-mode phonon dynamics and lasing in optomechanical cavities"}

\author{Raúl Ortiz}
\email{rortfer@ntc.upv.es}
\affiliation{Nanophotonics Technology Center, Universitat Politècnica de València, Building 8F, 46022 Valencia, Spain}

\author{Carlos Mas Arabí}
\author{Carles Milián}
\affiliation{Institut Universitari de Matemàtica Pura i Aplicada, Universitat Politècnica de València, 46022 Valencia, Spain}

\author{Alejandro Martínez}
\affiliation{Nanophotonics Technology Center, Universitat Politècnica de València, Building 8F, 46022 Valencia, Spain}

\date{\today}

\maketitle

\section{OM equations: normalization and steady-state}
\subsection{One optical mode coupled to one mechanical mode}
The optomechanical (OM) dynamical equations for one optical mode ($\tilde{a}$) coupled to one mechanical mode ($\tilde{x}$) are:
\begin{subequations}
\renewcommand{\theequation}{S1\alph{equation}}
\begin{align}
& \frac{d \tilde{a}}{d \tilde{t}}=\left(i(\tilde{\Delta}+\tilde{G} \tilde{x})-\frac{\tilde{\kappa}}{2}\right) \tilde{a}+\sqrt{\tilde{\kappa}_e} \tilde{S}_{\text{in}}, \\
& \frac{d^2 \tilde{x}}{d \tilde{t}^2}=- \tilde{\Omega}^2 \tilde{x}-\tilde{\Gamma} \frac{d \tilde{x}}{d \tilde{t}}+\frac{\hbar \tilde{G}}{\tilde{m}} |\tilde{a}|^2.
\end{align}
\label{1OM}
\end{subequations}

\begin{figure}
    \centering
    \includegraphics[width=0.9\columnwidth]{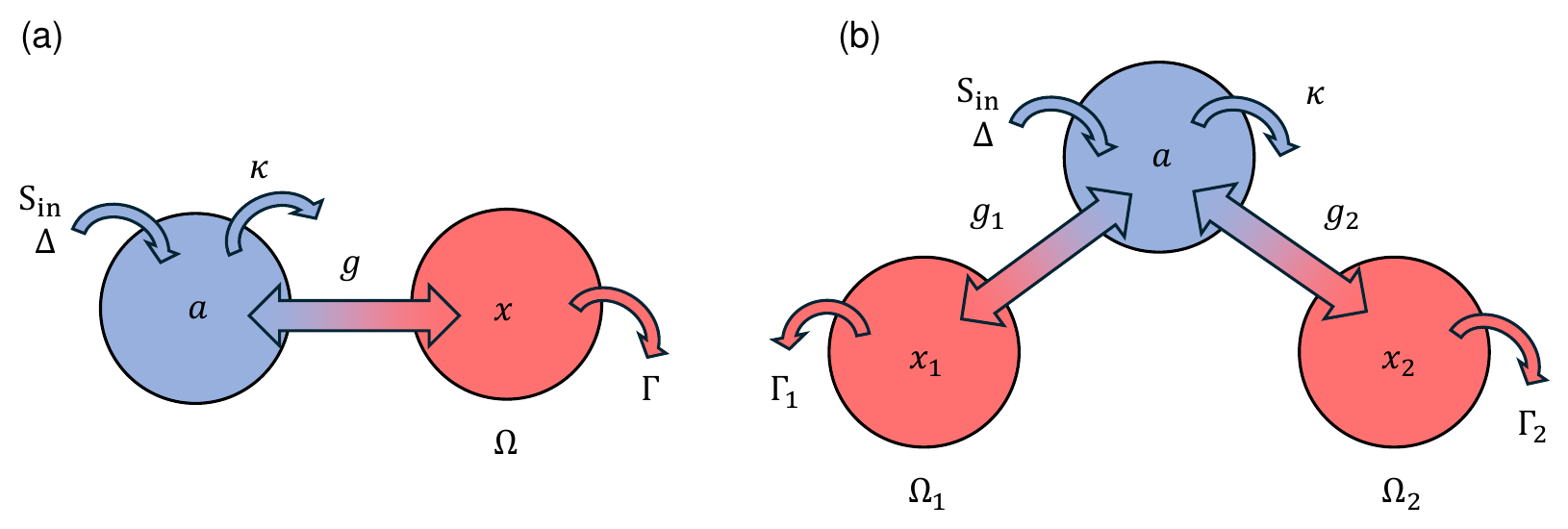}
    \caption{Schemes of the OM single-mode (a) and two-mode (b) interaction.}
    \label{balls1mode}
\end{figure}

Here, $\tilde{\Delta}=\tilde{\omega}_l-\tilde{\omega}_r$ is the detuning of the laser compared to the optical resonance, $\tilde{G}$ is the optical frequency shift per displacement unit, $\tilde{\kappa}$ is the overall optical loss and $\tilde{\kappa}_e$ represents the input coupling losses. $\tilde{S}_{in}$ is the input photon flux, $\tilde{m}$ is the mass of the mechanical oscillator and $\tilde{\Gamma}$ stands for the mechanical losses. Figure \ref{balls1mode}(a) illustrates the system described by Eqs. S\ref{1OM}(a),(b).
To study the stability of these equations, it is numerically convenient to normalize them:
\begin{subequations}
\renewcommand{\theequation}{S2\alph{equation}}
\begin{align}
& \frac{d a}{d \tau}=\left(i\left(\Delta+2 x\right)-1\right) a+S, \\
& \frac{d^2 x}{d \tau^2}=-\Omega^2 x - \Gamma \frac{d x}{d \tau} + \left|a\right|^2,
\end{align}
\label{1OM_N}
\end{subequations}
with $a$ and $x$ the optical and mechanical normalized modes. In this normalization we have taken $\tau=\dfrac{\tilde{\kappa}}{2}\tilde{t}$, $x=\dfrac{\tilde{G}\tilde{x}}{\tilde{\kappa}}$ and $a=\sqrt{\dfrac{4\tilde{G}^2\hbar}{\tilde{m}\tilde{\kappa}^3}}\tilde{a}$. On the other hand, $\Delta=\dfrac{2 \tilde{\Delta}}{\tilde{\kappa}}$, $\Omega^2=\dfrac{4 \tilde{\Omega}^2}{\tilde{\kappa}^2}$, $\Gamma=\dfrac{2}{\tilde{\kappa}} \tilde{\Gamma}$ and $S=\sqrt{\dfrac{16\tilde{G}^2\hbar\tilde{\kappa}_e}{\tilde{m}\tilde{\kappa}^5}}\tilde{S}_{in}$. By using these non-dimensional equations, the parameter values become more manageable when considering experimentally feasible ranges. \\
\indent 
To solve the steady-state solution we must take $\dfrac{d a}{d\tau}=\dfrac{d x}{d\tau}=\dfrac{d ^2x}{d\tau^2}=0$, so that $a(\tau)=\overline{a}$ and $x(\tau)=\overline{x}$. Now if we solve Eqs. S\ref{1OM_N}(a),(b) we obtain:
\begin{subequations}
\renewcommand{\theequation}{S3\alph{equation}}
\begin{align}
\overline{x} &= \frac{\left|\overline{a}\right|^2}{\Omega^2}, \\
\overline{a} &= \frac{-S}{i\left(\Delta+2 \overline{x}\right)-1}.
\end{align}
\label{1ax_Norm}
\end{subequations}
Then, if we substitute Eq. S\ref{1ax_Norm}(a) into Eq. S\ref{1ax_Norm}(b) we derive the characteristic polinomy:
\begin{equation}
\renewcommand{\theequation}{S4}
\frac{4}{\Omega^4}\left|\overline{a}\right|^6+\frac{4 \Delta}{\Omega^2}\left|\overline{a}\right|^4+\left(\Delta^2+1\right)\left|\overline{a}\right|^2-\left|S\right|^2=0.
\label{CHPol1}
\end{equation}
Equation S\ref{CHPol1} contains all the steady-state solution curves as a function of the different OM system parameters. The representation of the steady-state solution is shown for different examples in the main text in Figs. 2(I)-(IV). Now if we make $\frac{dS}{d|\overline{a}|^2}=0$, the upper and lower saddle node points for a red-detuned case have the following expressions: 
\begin{subequations}
\renewcommand{\theequation}{S5\alph{equation}}
\begin{align}
\left|\overline{a}\right|^2 &=  \Omega^2\left(-\frac{1}{3} \Delta \pm \sqrt{\left(\frac{\Delta^2}{36}-\frac{1}{12}\right)}\right), \\
\Delta &= -\frac{4\left|\overline{a}\right|^2}{\Omega^2} \pm \sqrt{\frac{4\left|\overline{a}\right|^4}{\Omega^4}-1}.
\end{align}
\label{SN}
\end{subequations}

From Eqs. S\ref{SN}(a),(b) we obtain the bistability condition $\Delta<-\sqrt{3}$.

\subsection{One optical mode coupled to two mechanical modes}
In this case we add a second mechanical mode ($\tilde{x}_{2}$) coupled to the same optical mode:
\begin{subequations}
\renewcommand{\theequation}{S6\alph{equation}}
\begin{align}
\frac{d \tilde{a}}{d \tilde{t}} &= \left(i\left(\tilde{\Delta} + \tilde{G}_1 \tilde{x}_1 + \tilde{G}_2 \tilde{x}_2\right) - \frac{\tilde{\kappa}}{2}\right) \tilde{a} + \sqrt{\tilde{\kappa}_e} \tilde{S}_{\text{in}}, \\
\frac{d^2 \tilde{x}_1}{d \tilde{t}^2} &= -\tilde{\Omega}_1^2 \tilde{x}_1 - \tilde{\Gamma}_1 \frac{d \tilde{x}_1}{d \tilde{t}} + \frac{\hbar \tilde{G}_1 |\tilde{a}|^2}{\tilde{m}_1}, \\
\frac{d^2 \tilde{x}_2}{d \tilde{t}^2} &= -\tilde{\Omega}_2^2 \tilde{x}_2 - \tilde{\Gamma}_2 \frac{d \tilde{x}_2}{d \tilde{t}} + \frac{\hbar \tilde{G}_2 |\tilde{a}|^2}{\tilde{m}_2}.
\end{align}
\end{subequations}
In principle, we assume that the second mechanical mode is different from the first mechanical mode to generally solve the problem. Figures \ref{balls1mode}(b) and 1(i) from the main text shows the scheme of the OM two-mode interaction between one optical and two mechanical modes.
Under these circumstances, the equations are normalized as follows:
\begin{subequations}
\renewcommand{\theequation}{S7\alph{equation}}
\begin{align}
\frac{d a}{d \tau} &= \left(i\left(\Delta + 2 \frac{x_{1}}{\gamma_2} + 2 \frac{x_{2}}{\gamma_1}\right) - 1\right) a + S, \\
\frac{d^2 x_{1}}{d \tau^2} &= -\Omega_{1}^2 x_{1} - \Gamma_{1} \frac{d x_{1}}{d \tau} + \left|a\right|^2, \\
\frac{d^2 x_{2}}{d \tau^2} &= -\Omega_{2}^2 x_{2} - \Gamma_{2} \frac{d x_{2}}{d \tau} + \left|a\right|^2.
\end{align}
\label{2OMNorm}
\end{subequations}
In this case, the transformations that we used are $\tau=\dfrac{\tilde{\kappa}}{2}\tilde{t}$, $x_{i}=\dfrac{\tilde{G}_i\gamma_j}{\tilde\kappa}\tilde{x}_i$,
$\gamma_i=\dfrac{4 \hbar \tilde{G}_{j}^2 }{\tilde{m}_j\tilde{\kappa}^3}$ $a=\sqrt{\gamma_1\gamma_2} \tilde{a}$, $\Delta=\dfrac{2 \tilde{\Delta}}{\tilde{\kappa}}$, $\Omega_{i}^2=\dfrac{4 \tilde{\Omega}{_i}^2}{\tilde{\kappa}^2}$, $\Gamma_{i}=\dfrac{2}{\tilde{\kappa}} \tilde{\Gamma}_{i}$ and $S=\dfrac{2}{\tilde\kappa}\sqrt{\gamma_1\gamma_2\kappa_e}\tilde{S}_{in}$. It should be noted that each mechanical mode is normalized using parameters from both mechanical modes. \\
\indent 
Although the normalized equations with two modes can be studied in their most general form, in this work, we focus on the particular case where $\gamma_1=\gamma_2\rightarrow\dfrac{\tilde{G}_{1}^2 }{\tilde{m}_1}=\dfrac{\tilde{G}_{2}^2 }{\tilde{m}_2}\rightarrow\tilde{\Omega}_1\tilde{g}_{0_1}^2=\tilde{\Omega}_2\tilde{g}_{0_2}^2$. This choice simplifies the numerical analysis without loss of generality, as both the methodology employed and the results obtained remain applicable to the general case, preserving the qualitative validity of the conclusions. Hence, the equations are:
\begin{subequations}
\renewcommand{\theequation}{S8\alph{equation}}
\begin{align}
\frac{d a}{d \tau} &= \left(i\left(\Delta + 2 x_{1} + 2 x_{2}\right) - 1\right) a + S, \\
\frac{d^2 x_{1}}{d \tau^2} &= -\Omega_{1}^2 x_{1} - \Gamma_{1} \frac{d x_{1}}{d \tau} + \left|a\right|^2, \\
\frac{d^2 x_{2}}{d \tau^2} &= -\Omega_{2}^2 x_{2} - \Gamma_{2} \frac{d x_{2}}{d \tau} + \left|a\right|^2.
\end{align}
\label{2OMNorm2}
\end{subequations}
The new normalizations are: $x_{i}=\dfrac{\tilde{G}_i}{\tilde\kappa}\tilde{x}_i$,
$\gamma=\dfrac{4 \hbar \tilde{G}_{i}^2 }{\tilde{m}_i\tilde{\kappa}^3}$ $a=\sqrt{\gamma} \tilde{a}$, and $S=\dfrac{2\sqrt{\gamma\kappa_e}}{\tilde\kappa}\tilde{S}_{in}$.
\\
\indent 
Conversely, the steady-state solutions are:
\begin{subequations}
\renewcommand{\theequation}{S9\alph{equation}}
\begin{align}
\overline{x}_{i} &= \frac{\left|\overline{a}\right|^2}{\Omega_{i}^2}, \\
\overline{a} &= \frac{-S}{i\left(\Delta + 2 \overline{x}_{1} + 2 \overline{x}_{2}\right) - 1}.
\end{align}
\label{Stat2OM}
\end{subequations}
The expressions in Eqs. S\ref{Stat2OM}(a),(b) are closely resembling to the single-mode ones (Eqs. S\ref{1ax_Norm}(a),(b)) but we have to take into account that the new normalizations of $a$ and $x_{i}$ are different.\\
\indent 
Now substituting Eq. S\ref{Stat2OM}(a) into Eq. S\ref{Stat2OM}(b) we derive the characteristic polinomy:

\begin{subequations}
\renewcommand{\theequation}{S10}
\begin{align}
4\left(\frac{1}{\Omega_{1}^4}+\frac{1}{\Omega_{2}^4}+\frac{2}{\Omega_{1}^2 \Omega_{2}^2}\right)\left|\overline{a}\right|^6+4 \Delta\left(\frac{1}{\Omega_{1}^2}+\frac{1}{\Omega_{2}^2}\right)\left|\overline{a}\right|^4 
+\left(\Delta^2+1\right)\left|\overline{a}\right|^2-\left|S\right|^2=0.
\end{align}
\end{subequations}

Following the same steps than in the single mechanical mode model, the saddle nodes have the following expressions:
\begin{subequations}
\renewcommand{\theequation}{S11\alph{equation}}
\begin{align}
\left|\overline{a}\right|^2 &= \left(\frac{\Omega_{1}^2 \Omega_{2}^2}{\Omega_{1}^2+\Omega_{2}^2}\right)\left[-\frac{1}{3} \Delta \pm \sqrt{\left(\frac{\Delta^2}{36}-\frac{1}{12}\right)}\right] ,\\
\Delta &= -\frac{4\left|\overline{a}\right|^2\left(\Omega_{1}^2+\Omega_{2}^2\right)}{\Omega_{1}^2 \Omega_{2}^2} \pm \sqrt{\frac{4\left|\overline{a}\right|^4\left(\Omega_{1}^2+\Omega_{2}^2\right)^2}{\left(\Omega_{1}^2 \Omega_{2}^2\right)^2}-1}.
\end{align}
\label{SN2OM}
\end{subequations}

Eqs. S\ref{SN2OM}(a),(b) are equivalent to the case of one mechanical mode (Eqs. S\ref{SN}(a),(b)) but taking the transformation $\Omega^2 \rightarrow \left(\dfrac{1}{\Omega_{1}^2}+\dfrac{1}{\Omega_{2}^2}\right)^{-1}$. It is also remarkable that we preserve the condition $\Delta<-\sqrt{3}$ for bistability.\\

\textbf{\underline{One mechanical mode limit case}}\\
\indent
We can simplify from the two-mode to the single-mode case by setting one mechanical frequency, $\Omega_{i}$, much larger than the other, $\Omega_{j}$.  This becomes apparent in Eqs. S\ref{Stat2OM}(a),(b) where for large $\Omega_{i}$, the contribution from $\overline{x}_{i}$ becomes negligible.\
\indent 
Nonetheless, when the two modes are degenerate (i.e., when they have identical mechanical properties), the two-mode equations can be reformulated to resemble the single-mode case.\\
\indent 
Let's first consider the normalized two-mode equations, Eqs. S\ref{2OMNorm2}(a)-(c). We define a new variable $X=x_{1}+x_{2}$. Hence, we consolidate the mechanical equations into a single equation:
\begin{subequations}
\renewcommand{\theequation}{S12\alph{equation}}
\begin{align}
    \frac{d a}{d \tau} &= \left(i\left(\Delta + 2 X\right) - 1\right) a + S ,\\
    \frac{d^2 X}{d \tau^2} &= -\Omega^2 X - \Gamma \frac{d X}{d \tau} + 2\left|a\right|^2.
\end{align}
\end{subequations}
Now, if we rewrite $2|a|^2=|A|^2$ and $\sqrt{2}S=C$ we finally reach the single-mode case:
\begin{subequations}
\renewcommand{\theequation}{S13\alph{equation}}
\begin{align}
\renewcommand{\theequation}{S13\alph{equation}}
    \frac{d A}{d \tau} &= \left(i\left(\Delta + 2 X\right) - 1\right) A + C ,\\
    \frac{d^2 X}{d \tau^2} &= -\Omega^2 X - \Gamma \frac{d X}{d \tau} + \left|A\right|^2.
\end{align}
\end{subequations}

\section{Stability analysis}
In the previous section we have obtained the steady-state solutions. Nevertheless, the behavior of the equations is quite more complex. For sufficiently large input powers ($S>>$), the steady-state solutions become unstable, leading to a range of nonlinear phenomena such as oscillations, period doublings, multistability, and chaos.\\
\indent 
The stability analysis of the equations identifies the bifurcations where stability changes occur \cite{NLDC}. This formalism involves considering the steady-state solutions with a time-dependent, sufficiently small perturbation. This approach leads to the calculation of a Jacobian matrix, whose eigenvalues provide information about the stability at each point. The necessary condition for stability is that all eigenvalues' real parts must be negative. If at least one of them becomes positive, the solution is no longer stable and will collapse to a stable attractor \cite{Stab_Bif}.

Conversely, if the real part of a pair of eigenvalues becomes positive and one is the complex conjugate of the other, while the real parts of the remaining eigenvalues stay negative, the system will exhibit oscillations. The transition of stability is always marked by a bifurcation; thus, identifying these points is essential for fully characterizing the system's stability. 
 
With this brief introduction, we can show the theoretical treatment of the stability analysis. First, we will derive the single mechanical mode model and then, we will develop the model for two mechanical modes. \\
\indent 
Let's consider this ansatz: $a(\tau) \approx \overline{a}+\delta a(\tau)$; $x(\tau) \approx \overline{x}+\delta x(\tau)$. We can take the Taylor series expansion of both variables with this general formula:
\begin{equation}
\renewcommand{\theequation}{S14}
\frac{d(\overline{V}+\delta V)}{d t}=f(\overline{V}+\delta V) \approx f(\overline{V})+\left(\frac{\partial f(V)}{\partial V_k}\right)_{V=\overline{V}} \delta V=\left(\frac{\partial f(V)}{\partial V_k}\right)_{V=\overline{V}} \delta V.
\label{Taylor}
\end{equation}

being $V$ either $a$ or $x$. Eq. S\ref{Taylor} shows that the time evolution of the optical and mechanical modes is proportional to the Jacobian matrix $\mathcal{J}=\left(\dfrac{\partial f(V)}{\partial V_k}\right)_{V=\overline{V}}$.\\
\indent 
With this principle, we can construct the Jacobian matrix of the OM equations. To linearize the equations, we define the vector $\Psi=\left(\delta a, \delta a^*, \delta x, \delta P\right)^T$ with $P=\dfrac{dx}{d\tau}$.\\
\indent 
Now, introducing the first ansatz into Eqs. S\ref{1OM}(a),(b) we can define the following equations:
\begin{subequations}
\renewcommand{\theequation}{S15\alph{equation}}
\begin{align}
f_1 &\equiv \frac{d \delta a}{d \tau}=\left(i\left(\Delta+2\left(\overline{x}+\delta x\right)-1\right) \delta a+2 i \overline{a} \delta x\right), \\
f_2 &\equiv \frac{d \delta a^*}{d \tau}=\left(-i\left(\Delta+2\left(\overline{x}+\delta x\right)-1\right) \delta a^*-2 i \overline{a}^* \delta x\right) ,\\
f_3 &\equiv \frac{d \delta x}{d \tau}=\delta P ,\\
f_4 &\equiv \frac{d \delta P}{d \tau}=-\Omega^2 \delta x-\Gamma \delta P+\left(\overline{a} \delta a^*+\overline{a}^* \delta a\right).
\end{align}
\end{subequations}
Naturally, the time evolution of the steady-state solutions is null, i.e., $\dfrac{d \overline{a}}{d \tau}=\dfrac{d \overline{x}}{d \tau}=0$, and we have neglected the ${O}^2$ term $|\delta a|^2$. The $\mathcal{J}$ matrix is calculated as follows:
\begin{equation}
\renewcommand{\theequation}{S16}
\label{eq:matriz_derivadas}
\begin{aligned}
\renewcommand{\arraystretch}{1.5}
\mathcal{J}=
\begin{pmatrix}
\frac{\partial f_1}{\partial \delta a} & \frac{\partial f_1}{\partial \delta a^*} & \frac{\partial f_1}{\partial \delta x} & \frac{\partial f_1}{\partial \delta P} \\
\frac{\partial f_2}{\partial \delta a} & \frac{\partial f_2}{\partial \delta a^*} & \frac{\partial f_2}{\partial \delta x} & \frac{\partial f_2}{\partial \delta P} \\
\frac{\partial f_3}{\partial \delta a} & \frac{\partial f_3}{\partial \delta a^*} & \frac{\partial f_3}{\partial \delta x} & \frac{\partial f_3}{\partial \delta P} \\
\frac{\partial f_4}{\partial \delta a} & \frac{\partial f_4}{\partial \delta a^*} & \frac{\partial f_4}{\partial \delta x} & \frac{\partial f_4}{\partial \delta P}
\end{pmatrix}
\end{aligned}
\end{equation}

Hence, the time evolution of the optical and mechanical modes is $\dfrac{d}{d \tau}\Psi=\mathcal{J}\Psi$:
\begin{equation}
\renewcommand{\theequation}{S17}
\label{eq:ecuacion_diferencial}
\frac{d}{d \tau}\begin{pmatrix}
\delta a \\
\delta a^* \\
\delta x \\
\delta P
\end{pmatrix} = \begin{pmatrix}
i\left(\Delta+2 \overline{x}\right)-1 & 0 & 2 i \overline{a} & 0 \\
0 & -i\left(\Delta+2 \overline{x}\right)-1 & -2 i \overline{a}^* & 0 \\
0 & 0 & 0 & 1 \\
\overline{a}^* & \overline{a} & -\Omega^2 & -\Gamma
\end{pmatrix}\begin{pmatrix}
\delta a \\
\delta a^* \\
\delta x \\
\delta P
\end{pmatrix}
\end{equation}
For the case of two mechanical modes, the procedure is similar. We assume the same ansatz:  $a(\tau) \approx \overline{a}+\delta a(\tau) ;\; x_{1}(\tau) \approx \overline{x}_{1}+\delta x_{1}(\tau) ;\; x_{2}(\tau) \approx \overline{x}_{2}+\delta x_{2}(\tau)$. In this case, the equations for the $\mathcal{J}$ matrix are:
\begin{subequations}
\renewcommand{\theequation}{S18\alph{equation}}
\label{eq:sistema_ecuaciones}
\begin{align}
& f_1 \equiv \frac{d \delta a}{d \tau}= \left(i\left(\Delta+2\left(\overline{x}_{1}+\delta x_{1}\right)+2\left(\overline{x}_{2}+\delta x_{2}\right)\right)-1\right) \delta a+2 i \overline{a}\left(\delta x_{1}+\delta x_{2}\right), \\
& f_2 \equiv \frac{d \delta a^*}{d \tau}= \left(-i\left(\Delta+2\left(\overline{x}_{1}+\delta x_{1}\right)+2\left(\overline{x}_{2}+\delta x_{2}\right)\right)-1\right) \delta a^*-2 i \overline{a}^*{ }\left(\delta x_{1}+\delta x_{2}\right) ,\\
& f_3 \equiv \frac{d \delta x_{1}}{d \tau}= \delta P_{1}, \\
& f_4 \equiv \frac{d \delta P_{1}}{d \tau}= -\Omega_{1}^2 \delta x_{1}-\Gamma_{1} \delta P_{1}+\left(\overline{a} \delta a^*+\overline{a}{ }^* \delta a\right), \\
& f_5 \equiv \frac{d \delta x_{2}}{d \tau}= \delta P_{2} ,\\
& f_6 \equiv \frac{d \delta P_{2}}{d \tau}= -\Omega_{2}^2 \delta x_{2}-\Gamma_{2} \delta P_{2}+\left(\overline{a} \delta a^*+\overline{a}^* \delta a\right).
\end{align}
\end{subequations}
Here, the vector of parameters is $\Psi=\left(\delta a, \delta a^*, \delta x_{1}, \delta P_{1}, \delta x_{2}, \delta P_{2}\right)^T$, with $P_{{1}}=\dfrac{dx_{{1}}}{d\tau}$ and $P_{{2}}=\dfrac{dx_{{2}}}{d\tau}$. Finally, the system of time evolution equations are:

\begin{equation}
\renewcommand{\theequation}{S19}
\label{eq:ecuacion_diferencial_vectorial}
\begin{aligned}
\small 
\frac{d}{d \tau}
\begin{pmatrix}
\delta a \\
\delta a^* \\
\delta x_{1} \\
\delta P_{1} \\
\delta x_{2} \\
\delta P_{2}
\end{pmatrix} &=
\begin{pmatrix}
i\left(\Delta+2 \overline{x}_{1}+2 \overline{x}_{2}\right)-1 & 0 & 2 i \overline{a} & 0 & 2 i \overline{a} & 0 \\
0 & -i\left(\Delta+2 \overline{x}_{1}+2 \overline{x}_{2}\right)-1 & -2 i \overline{a}^* & 0 & -2 i \overline{a}^* & 0 \\
0 & 0 & 0 & 1 & 0 & 0 \\
\overline{a}^* & \overline{a} & -\Omega_{1}^2 & -\Gamma_{1} & 0 & 0 \\
0 & 0 & 0 & 0 & 0 & 1 \\
\overline{a}^* & \overline{a} & 0 & 0 & -\Omega_{2}^2 & -\Gamma_{2} 
\end{pmatrix}
\begin{pmatrix}
\delta a \\
\delta a^* \\
\delta x_{1} \\
\delta P_{1} \\
\delta x_{2} \\
\delta P_{2}
\end{pmatrix}
\end{aligned}
\end{equation}

The stability of the solutions can be investigated by evaluating the eigenvalues of the Jacobian matrices, as we explained at the beginning of this subsection. Equations S\ref{eq:both_equations}(a),(b) show the solutions of the eigenvalues ($|\mathcal{J}-\lambda I|=0$) for one (a) and two (b) mechanical modes.
$I$ accounts for the associated 4x4 (single-mode) and 6x6 (two-modes) rank identity matrices.

\begin{subequations}
\renewcommand{\theequation}{S20\alph{equation}}
\label{eq:both_equations}
\small
\begin{align}
& \quad \lambda^4 + \lambda^3 (2+\Gamma ) + \lambda^2 (1+2\Gamma+(2\overline{x}+\Delta)^2+\Omega^2) + \lambda (\Gamma+\Gamma(2\overline{x}+\Delta)^2+2\Omega^2 ) + \Omega^2 + (2\overline{x}+\Delta)^2\Omega^2 + 4\overline{a}(2\overline{x}+\Delta)\overline{a}^* , && \\[7pt]
& \quad \lambda^6 +\notag \\
&  \quad \lambda^5 (2+\Gamma_{1}+\Gamma_{2}) +\notag \\
& \quad \lambda^4 \big(1+2\Gamma_{1}+(2+\Gamma_{1})\Gamma_{2}+(2(\overline{x}_{1}+\overline{x}_{2})+\Delta)^2+{\Omega_{1}}^2+{\Omega_{2}}^2 \big) +\notag \\ &  \quad \lambda^3 \big(\Gamma_{1}(1+(2(\overline{x}_{1}+x_{2})+\Delta)^2) +2{\Omega_{1}}^2+\Gamma_{2}(1+2\Gamma_{1}+(2(\overline{x}_{1}+\overline{x}_{2})+\Delta)^2+{\Omega_{1}}^2 )+(2+\Gamma_{1}){\Omega_{2}}^2 \big) +\notag \\
&  \quad \lambda^2 \big((1+(2({x}_{1}+\overline{x}_{2})+\Delta)^2){\Omega_{1}}^2+\Gamma_{2}(\Gamma_{1}(1+(2(\overline{x}_{1}+\overline{x}_{2})+\Delta)^2)+2{\Omega_{1}}^2 ) +(1+2\Gamma_{1}+(2(\overline{x}_{1}+\overline{x}_{2})+\Delta)^2+{\Omega_{1}}^2){\Omega_{2}}^2+\notag \\
& \;\;\quad\;\;8\overline{a}(2(\overline{x}_{1}+\overline{x}_{2})+\Delta)\overline{a}^* \big) + \notag \\
& \quad \lambda \big(\Gamma_{2}(1+(2(\overline{x}_{1}+\overline{x}_{2})+\Delta)^2){\Omega_{1}}^2+(\Gamma_{1}(1+(2(\overline{x}_{1}+\overline{x}_{2})+\Delta)^2)+2{\Omega_{1}}^2 ){\Omega_{2}}^2 +4\overline{a}(\Gamma_{1}+\Gamma_{2})(2(\overline{x}_{1}+\overline{x}_{2})+\Delta)\overline{a}^* \big)+ \notag \\
&  \quad(1+(2(\overline{x}_{1}+\overline{x}_{2})+\Delta)^2){\Omega_{1}}^2{\Omega_{2}}^2+4\overline{a}(2(\overline{x}_{1}+\overline{x}_{2})+\Delta)({\Omega_{1}}^2+{\Omega_{2}}^2 )\overline{a}^* .
\end{align}
\end{subequations}

Due to the complexity of both polinomial equations, it is necessary to solve them numerically. For that purpose we use numerical time evolution simulations, and AUTO \cite{Auto}, a continuation software that efficiently tracks bifurcations and periodic solutions by following steady-state branches as parameters are varied. Unlike direct time-evolution simulations, AUTO allows for the precise identification of bifurcation points and stable periodic solutions without requiring long simulation times, providing a more efficient and robust analysis of the stability of the system.
\section{Normalized to SI parameters}
The experimentally achievable parameters used in Fig. 2(Fig. 3) of the main text, along with their normalized values, are shown in Table I(Table II). \\
\indent
The values of $\tilde\kappa$, $\tilde\kappa_e$, $\tilde{g}_{0_{1}}$, $\tilde{g}_{0_{2}}$, $\tilde{m}_1$ and $\tilde{m}_2$ have been arbitrarily chosen, as they are not uniquely determined by the selected set of normalized parameters. This selection considers the condition $\tilde{\Omega}_1 \tilde{g}_{0_{1}}^2 = \tilde{\Omega}_2 \tilde{g}_{0_{2}}^2$ (see transformations in Section I). Nevertheless, these values remain realistic and experimentally feasible.

\begin{table*}[t]
    \centering
    \begin{ruledtabular}
        \begin{tabular}{cccccccccccccccc}
            \textbf{Normalized} & $\Omega_{1}$ & $\Omega_{2}$ & $\Gamma_{1}$ & $\Gamma_{2}$ & $\Delta$ & $S$ \\ 
            \textbf{Values}     & $2$ & $3$ & $10^{-5}$ & $10^{-5}$ & $1$ & $0.03$ \\ 
            \hline
            \textbf{SI} & $\tilde\Omega_1$ (GHz) & $\tilde\Omega_2$ (GHz)& $\tilde\Gamma_1$ (kHz)& $\tilde\Gamma_2$ (kHz)& $\tilde\Delta$ (GHz)& $\tilde{S}$ ($s^{-1/2}$)& $\tilde{P}$ (nW) \\ 
            \textbf{Values}      & $2 \pi \times 2$ & $2 \pi \times 3$ &  $2 \pi \times 10$ & $2 \pi \times 10$ & $2 \pi$ & $5.3 \times 10^5$ & $36$ \\ 
            & $\tilde\kappa$ (GHz)& $\tilde\kappa_e$ (GHz)& $\tilde{g}_{0_1}$ (MHz)& $\tilde{g}_{0_2}$ (MHz)& $\tilde{m}_1$ (kg)& $\tilde{m}_2$ (kg)& \\ 
            & $2 \pi \times 2 $ & $2 \pi \times 1.5$ & $2\pi \times2.60$ & $2\pi \times2.12$ & $10^{-16}$ & $10^{-16}$ & \\ 
        \end{tabular}
    \end{ruledtabular}
    \caption{Feasible normalized and SI OM parameters corresponding to Fig. 2.}
    \label{tab:parameters}
\end{table*}

\begin{table*}[t]
    \centering
    \begin{ruledtabular}
        \begin{tabular}{cccccccccccccccc}
            \textbf{Normalized} & $\Omega_{1}$ & $\Omega_{2}$ & $\Gamma_{1}$ & $\Gamma_{2}$ & $\Delta$ & $S$ \\ 
            \textbf{Values}     & $0.1$ & $1.5$ & $10^{-3}$ & $10^{-3}$ & $2$ & $0.15$ \\ 
            \hline
            \textbf{SI} & $\tilde\Omega_1$ (GHz) & $\tilde\Omega_2$ (GHz)& $\tilde\Gamma_1$ (MHz)& $\tilde\Gamma_2$ (MHz)& $\tilde\Delta$ (GHz)& $\tilde{S}$ ($s^{-1/2}$)& $\tilde{P}$ ($\mu$W) \\ 
            \textbf{Values}      & $2 \pi \times 0.1$ & $2 \pi \times 1.5$ &  $2 \pi$ & $2 \pi$ & $2 \pi\times 2$ & $1.7 \times 10^7$ & $36$ \\ 
            & $\tilde\kappa$ (GHz)& $\tilde\kappa_e$ (GHz)& $\tilde{g}_{0_1}$ (MHz)& $\tilde{g}_{0_2}$ (MHz)& $\tilde{m}_1$ (kg)& $\tilde{m}_2$ (kg)& \\ 
            & $2 \pi \times 2 $ & $2 \pi \times 1.5$ & $2\pi \times1.84$ & $2\pi \times0.47$ & $10^{-16}$ & $10^{-16}$ & \\ 
        \end{tabular}
    \end{ruledtabular}
    \caption{Feasible normalized and SI OM parameters corresponding to Fig. 3.}
    \label{tab:parameters}
\end{table*}

\section{Threshold analysis for dual-mode lasing via parameter sweeps}
\begin{figure*}[t]
    \centering
    \includegraphics[width=\textwidth]{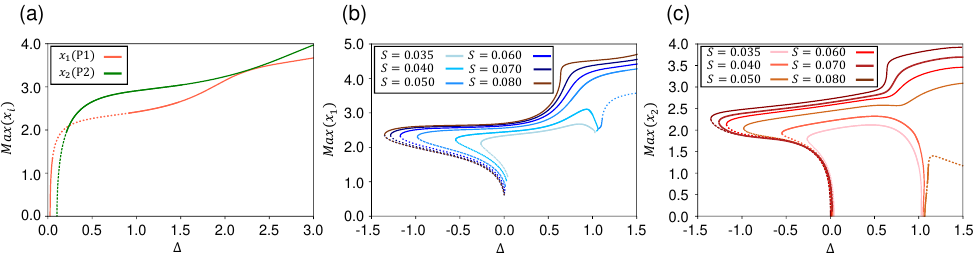}
    \caption{Amplitude of mechanical oscillations as a function of $\Delta$ for the case shown in Fig. 2.  
(a) Maximum amplitude of the mechanical oscillations for the $P_1$ and $P_2$ periodic branches.  
(b),(c) Corresponding results for the dual-mode lasing branch $P_{DL}$, using the same study as in (a) for $x_1$ and $x_2$, respectively.}
    \label{fig:Deltas}
\end{figure*}

\begin{figure*}[t]
    \centering
    \includegraphics[width=0.9\textwidth]{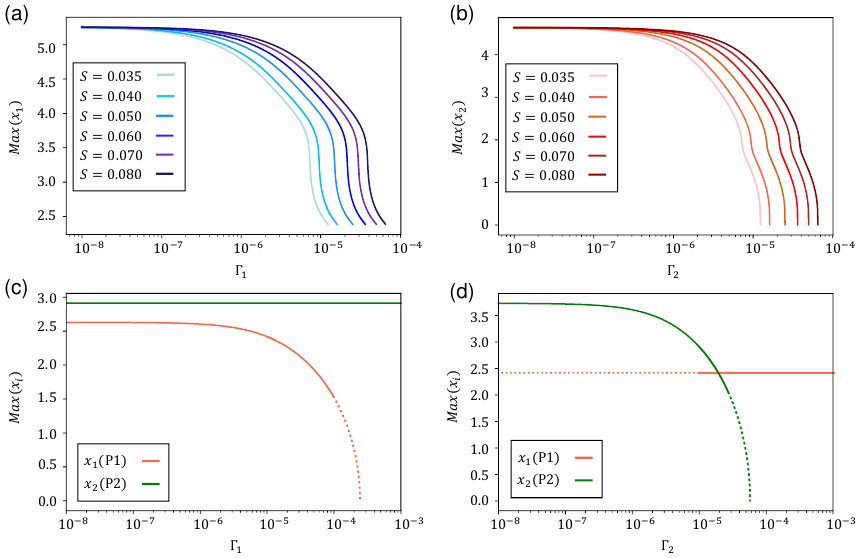}
    \caption{Maximum amplitude of the mechanical oscillations as a function of $\Gamma_{1}$ and $\Gamma_{2}$ for the case presented in Fig. 2. (a),(b) Sweeps over $\Gamma_{1}$ and $\Gamma_{2}$, respectively, for different values of $S$ corresponding to the $P_{DL}$ branch solution in Fig. 2(a). (c),(d) Sweeps over $\Gamma_{1}$ and $\Gamma_{2}$, respectively, for the periodic solution branches $P_1$ and $P_2$, also shown in Fig. 2(a).}
    \label{fig:Gammas}
\end{figure*}

\begin{figure*}[t!]
    \centering
    \includegraphics[width=0.6\textwidth]{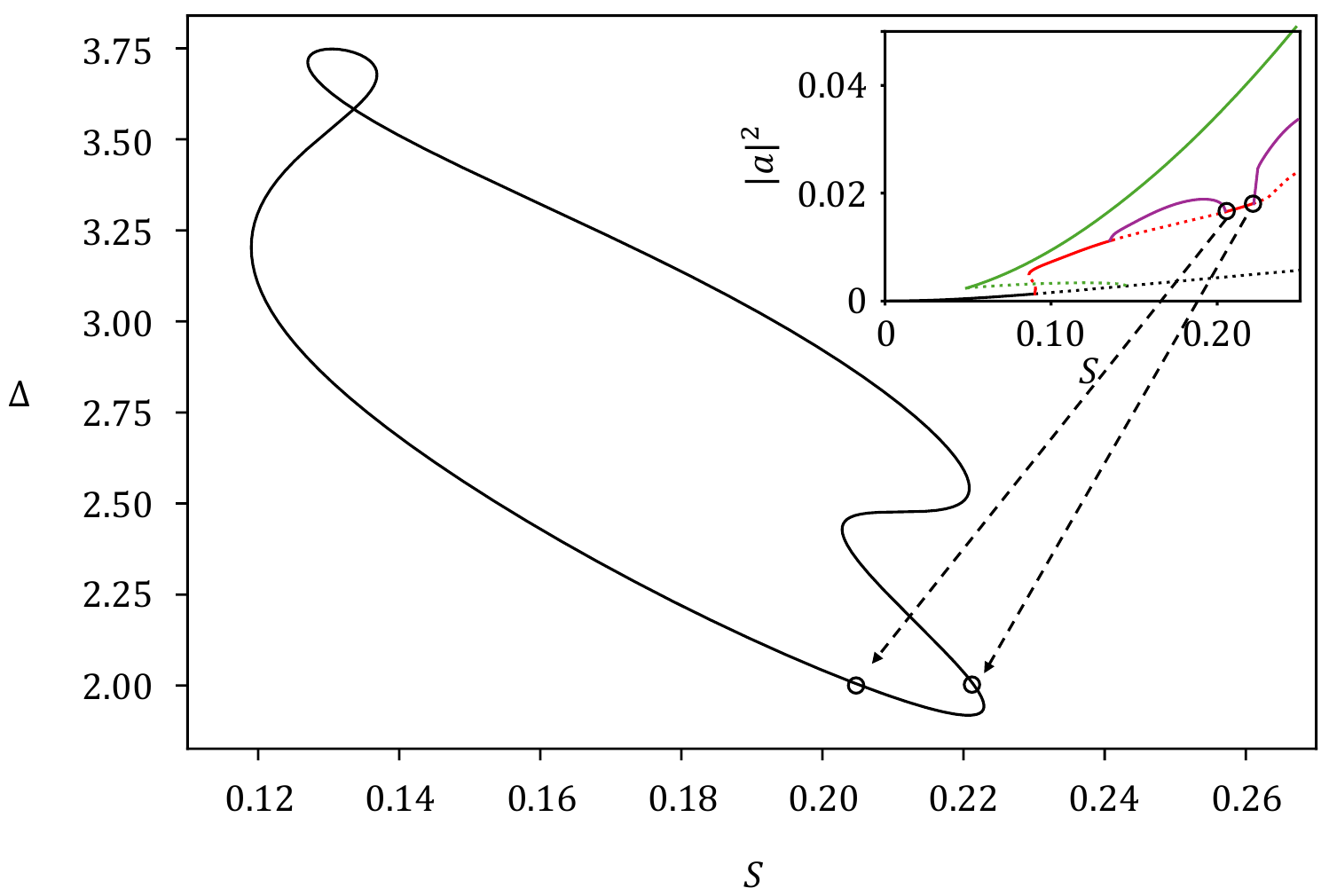}
    \caption{Sweep of the TR branch over $\Delta$ and $S$, corresponding to Fig. 3. Inset: Fig. 3(a) of the main text highlighting the TR bifurcations.}
    \label{fig:TR}
\end{figure*}

In this section, we explore the thresholds for dual-phonon lasing by tracking the continuation of the bifurcations (PD and TR) that lead to this solution from Figs. 2 and 3 across different system parameters. This analysis demonstrates the robustness of the solutions beyond specific parameter values.\\
\indent 
The dual-phonon lasing branch remains robust at higher values of $S$ as shown in Fig. 2. However, it is particularly interesting to examine its stability when varying different system parameters. In particular, Fig. \ref{fig:Deltas}(a) illustrates the oscillation amplitudes of both mechanical modes as a function of $\Delta$ for a fixed value of $S=0.05$ in $P_1$ and $P_2$ (see Fig. 2). As shown, comparable amplitudes for the two mechanical modes are observed only within a limited range of $\Delta$. However, since they belong to the $P_1$ and $P_2$ curves, which represent different solution branches, they do not coexist, meaning that only one of the mechanical modes can be enhanced at a given time. Notably, for this specific value of $S$, the oscillations become negligible for $\Delta < 0$. It is important to point out that $x_{{1}}$ in $P_2$ and $x_{{2}}$ in $P_1$ are not displayed, as their amplitudes remain negligible throughout.\\
Conversely, Figs. \ref{fig:Deltas}(b) and (c) show the amplitudes of $x_{{1}}$ and $x_{{2}}$, respectively, for varying values of $S$ in the dual-phonon lasing branch, $P_{DL}$. As $S$ increases, the curves broaden, encompassing a wider range of detunings over which both modes oscillate with comparable amplitudes. 
\\
\indent
A similar analysis for $\Gamma_{{1}}$ and $\Gamma_{{2}}$ is presented in Fig. \ref{fig:Gammas}. Figures \ref{fig:Gammas}(a) and (b) illustrate parameter sweeps over $\Gamma_{1}$ and $\Gamma_{2}$, respectively, for various values of $S$. Notably, for sufficiently low mechanical losses, the oscillation amplitudes converge to a specific value regardless of $S$. In contrast, at higher mechanical losses, the amplitudes decrease, with the rate of decrease depending on the pump strength $S$. Specifically, higher pump values sustain large oscillation amplitudes over a broader range of losses, while lower pump values result in a steeper decline in oscillation amplitude as losses increase. Figures \ref{fig:Gammas}(c) and (d) focus on the periodic solutions of $P_1$ and $P_2$ at $S=0.05$. As expected, for $P_1$ ($P_2$), only the first mechanical mode $x_{1}$ (second mechanical mode $x_{2}$) exhibits dominant oscillations. In particular, Fig. \ref{fig:Gammas}(c) shows that as $\Gamma_{1}$ increases, the amplitude of $x_{1}$ decreases, while $x_{2}$ remains unaffected. This behavior is consistent with the dynamics of the $P_2$ branch, where $x_{2}$ predominates, and changes in the losses of $x_{1}$ have minimal impact due to the highly attenuated nature of the first mechanical mode in this regime. In Fig. \ref{fig:Gammas}(d) we can see the same behavior for $x_{2}$. Notably, in this case, the amplitude of $x_{1}$ in $P_1$ remains unaffected whereas the second mode amplitude gets modified by $\Gamma_{2}$.
\\ 
\indent
On the other hand, torus bifurcations (TR) occur more frequently within the studied parameter range. This bifurcation leads to dual-phonon lasing in distinct regions, with each mode oscillating near its natural frequency. Figure \ref{fig:TR} shows the threshold curve of a TR bifurcation with respect to $S$ and $\Delta$ for the case presented in Fig. 3 and depicted in the inset of Fig. \ref{fig:TR}(c). As shown in the inset, the two consecutive TR bifurcations correspond to the same curve. It is also important to note that this curve does not delimit the dual-phonon lasing from other solutions, as additional bifurcations may appear both inside and outside the delimited region; it simply marks the threshold at which a specific TR bifurcation occurs.
\\
\indent
We have also performed calculations by sweeping the mechanical frequencies ($\Omega_{1}$ and $\Omega_{2}$). Unlike other parameters such as $\Delta$ or $\Gamma_i$, where bifurcations shift smoothly with continuous variations, sweeping $\Omega_{1}$ and $\Omega_{2}$ does not allow for infinitesimal continuation of bifurcations. Instead, bifurcations like period doubling (PD) and torus (TR) still occur for different $\Omega_i$ configurations, but their precise locations are not easily predictable, requiring an independent search for each case.\\
\indent
Notably, we have discovered that if the new comb lines generated by the PD bifurcation of the first lasing mode coincide with the natural frequency of the second mechanical mode, this can destabilize the original lasing branch. As a result, the period doubling bifurcation gives rise to a new stable branch corresponding to dual-mode lasing. This finding reveals a direct mechanism through which the PD bifurcation can mediate energy transfer between mechanical modes, enabling simultaneous lasing under specific conditions.

\vfill